\documentclass[12pt]{article}
\usepackage{psfig,epsf}
\usepackage{a4wide,amssymb,graphicx}
\usepackage{cite}

\begin{document}

\title{Mass dependence of inclusive nuclear $\phi$
photoproduction}

\maketitle

\begin{center}
{\large{D.~Cabrera$^1$, L.~Roca$^1$, E.~Oset$^1$,
 H.~Toki$^2$ and M.~J.~Vicente
Vacas$^1$}}

\vspace{.5cm}
{\it $^1$ Departamento de F\'{\i}sica Te\'orica and IFIC
 Centro Mixto Universidad de Valencia-CSIC\\
 Institutos de Investigaci\'on de Paterna, Apdo. correos 22085,
 46071, Valencia, Spain}    

\vspace{.5cm}
{\it $^2$Research Center for Nuclear Physics (RCNP), Osaka University,\\
Ibaraki, Osaka 567-0047, Japan
}  
\end{center}

 \begin{abstract} 

Based on a prior determination of the $\phi$ selfenergy in a
nuclear medium we perform a theoretical study of inclusive
$\phi$ photoproduction in nuclei, looking at the $A$
dependence of the cross sections for different $\phi$
momenta. We find sizeable reductions  in the nuclear cross
sections with respect to the elementary one, using a $\phi$
selfenergy which implies a width about six times the
free one at normal nuclear density. The calculations
are done to match the set up for an ongoing experiment at
{\it SPring8/Osaka} which should provide valuable information
on the renormalization of the $\phi$ properties in nuclei. 

\end{abstract}

\section{Introduction}

The renormalization of the properties of the vector mesons in
nuclear matter has received much attention, particularly the
$\rho$ meson, (see \cite{wambach} for a review). The $\phi$
meson has received comparatively less attention, but it turns
out to be a well suited tool to understand the dynamics of
the vector mesons in matter since the changes of some of the
$\phi$ properties are comparatively larger than those of the
$\rho$ meson, what would make its experimental observation
in principle easier. 
The change of the $\phi$ properties in the medium at finite temperature and/or
density has been studied in several
approaches like effective Lagrangians 
\cite{Blaizot:1991af,klingl2,Kuwabara:1995ms,Song:1996gw,Bhattacharyya:1997kx,klingl}
and QCD sum rules
\cite{Hatsuda:1991ez,Asakawa:1994tp,Zschocke:2002mn,Kampfer:2002pj}.
Particularly, the $\phi$ width modification in matter has
been subject of study in dropping meson mass scenarios
\cite{Bhattacharyya:1997kx,Ko:tp,Shuryak:1992yy,Lissauer:1991fr,Panda:1993ik,Ko:1994id},
as a result of collisional broadening through $\phi$-baryon \cite{Smith:1998xu}
or $\phi$-meson \cite{Alvarez-Ruso:2002ib,Alvarez-Ruso:wx}
 scattering  processes and by
modifying the $\bar{K}K$ decay channel by means of in-medium kaon selfenergies
\cite{klingl2,oset}. The majority of these works point at a
sizeable renormalization of the $\phi$ width and a small mass shift.

Indeed, in \cite{klingl,klingl2} it was
found that the $\phi$ width at normal nuclear matter density,
$\rho_0$, was about one order of magnitude larger than the
free width. Particularly, the results of \cite{klingl2} were based on the use of
a $K^-$ selfenergy in nuclear matter which accounted for Pauli blocking correction
in the intermediate $K^- N$ sates in the scattering equation. Further studies
considering the selfconsistency of the $\bar{K} N$ in-medium scattering equation
were done in \cite{lutz,ramos} to determine the $K^-$ selfenergy. Following
those results a
reanalysis of the $\phi$ width in the medium was done in
\cite{oset}, resulting in values around $22\,\textrm{MeV}$ at
$\rho=\rho_0$. A more recent evaluation, using the $K^-$
selfenergy of \cite{ramos} and improving on some
approximations of \cite{oset} was done in \cite{dani},
resulting in a width of about $28\,\textrm{MeV}$ at
$\rho=\rho_0$ and a small shift in the mass of about
$-6\,\textrm{MeV}$. 

Several proposals to test the $\phi$ properties in a nuclear medium have been
done, for instance, regarding the detection of its decay products in $A-A$ and
$p-A$ collisions \cite{Pal:2002aw,Kampfer:2002pj}. Particularly a method to analyse the 
invariant mass spectra of the emitted charged kaon pairs was studied in
\cite{Bhalerao:1997sj}.
Having into account that the calculations in \cite{klingl,klingl2,oset,dani}
were done for a $\phi$ at rest in the nucleus, it has been suggested to
determine this large width using reactions like $\pi^- p \to \phi n$ in nuclei
\cite{klingl2} and $\phi$ photoproduction in nuclei \cite{toki}, assuming the
$\phi$ to emerge with a small momentum. The latter proposal is possible 
by means of the elementary
$\gamma N\to\phi N$ reaction with the $\phi$ going in the
backward direction and the help of the Fermi motion of the
nucleons. The small cross section for these particular events
makes this experiment difficult and in addition it was shown
in \cite{mosel} that the distribution of the $K^+K^-$ pairs
from the Coulomb interaction in heavy nuclei removed the
changes in the $\phi$ width in the $K^+K^-$  invariant mass
that one might expect from the changes of the $\phi$ width in
the medium. Another possibility, not previously considered,
to investigate the changes of the $\phi$ width is to look for
$\phi$ loss of flux in $\phi$ nuclear photoproduction. The
$A$ dependence in this loss is related to the $\phi$ width in
the medium. The drawback is that in this experiment the
largest fraction of the $\phi$ comes out with a momentum of
the order of $1500\,\textrm{MeV/c}$ in the ongoing experiment
at {\it SPring8/Osaka} \cite{ishikawa}. The combination of
experiment and theoretical models, which can make the
extrapolation of $\phi$ at rest to $\phi$ with finite
momenta, can thus make the experiment useful to learn about 
$\phi$ nuclear properties.

With this aim, we present here a theoretical calculation of
the $A$ dependence of the cross section for $\phi$
photoproduction in nuclei which could test the models of
\cite{klingl,klingl2,oset,dani}. Our work will be based on
the models for the $\phi$ selfenergy of
Oset-Ramos~(OR)~\cite{oset} and
Cabrera-Vicente~(CV)~\cite{dani},
after a proper extension to account for the finite momentum.


\section{$\phi$ photoproduction cross section}

Let $\Pi(p,\rho)$ be the $\phi$ selfenergy in a nuclear
medium as a function of its momentum, $p$, and the nuclear
density, $\rho$. We have

\begin{equation}
\frac{\Pi}{2\omega}\equiv V_{\textrm{opt}}
= \ \textrm{Re} {V_{\textrm{opt}}} \ + \ i \, 
\textrm{Im} {V_{\textrm{opt}}} \ ,
\end{equation}

\noindent
and hence
\begin{equation}
\frac{\Gamma}{2}=- \textrm{Im}\frac{\Pi}{2\omega} \quad ; \qquad
\Gamma=-\frac{\textrm{Im}\Pi}{\omega}\equiv\frac{dP}{dt} \ ,
\end{equation}

\noindent 
where $\omega$ is the $\phi$ energy and $P$ is the
$\phi$ decay probability, including nuclear quasielastic and
absorption channels. Hence, we have for the probability of
loss of flux per unit length

\begin{equation}
\frac{dP}{dl}=\frac{dP}{v\,dt}=\frac{dP}{\frac{p}{\omega}dt}
=-\frac{\textrm{Im}\Pi}{p} \ .
\end{equation}

The nuclear cross section for inclusive $\phi$ photoproduction
will then be

\begin{equation}
\frac{d\sigma_A}{d\Omega}=\int d^3\vec{r}\rho(r)
\frac{d\sigma}{d\Omega}
\ e^{ -\int_0^{\infty}
dl \frac{-1}{p}\textrm{{\scriptsize Im}}\Pi(p,\rho(r'))}
\label{eq:eikonal1}
\end{equation}

\noindent  where $\frac{d\sigma}{d\Omega}$ and
$\frac{d\sigma_A}{d\Omega}$ are the elementary and nuclear
differential cross section and
$\vec{r}\,'=\vec{r}+l\frac{\vec p}{|\vec p|}$ with  $\vec{r}$
the $\phi$ production point inside the nucleus. The
exponential factor in Eq.~(\ref{eq:eikonal1}) represents the
survival probability of the $\phi$ meson in its way out of
the nucleus. If the $\phi$ did not get absorbed inside the
nucleus then we would get the typical result for an
electromagnetic reaction 
$\frac{d\sigma_A}{d\Omega}=A\frac{d\sigma}{d\Omega}$. 
Eq.~(\ref{eq:eikonal1}) relies in the eikonal approximation
which is accurate for the large $\phi$ momentum
involved in the process. In practice one might expect small
corrections to the formula, even if the $\phi$ did not decay,
from two sources:

i) Distortion of the $\phi$ trajectory because of the real
part of the potential

ii) Change of direction and energy of the $\phi$ in quasielastic
collisions $\phi N\to\phi'N'$.

\noindent 
The first effect should be negligible since the
real part of the potential is so weak that only modifies the
mass of the $\phi$ in about $6\,\textrm{MeV}$
\cite{klingl,dani}. The second effect should be very weak
too, account taken of  the lack of direct coupling $\phi NN$
because of the OZI rule. On the other hand, the effect from
this source would simply lead to a slight change in the
$\phi$ direction  but not the disappearance of the $\phi$.
This simply means that collecting the $\phi$ in a narrow cone
along the forward direction, where practically all the $\phi$
go, both because of the kinematics of the lab variables and the
extremely forward direction of the $\gamma N \to\phi N$ cross
section \cite{clas}, would guarantee that the $\phi$
undergoing quasielastic collisions in the nucleus are
accounted for in the nuclear cross section.
In order to adjust to this experiment we should not remove
theoretically the events in which there are $\phi$
quasielastic collisions. This will be done by including in
$\textrm{Im}\Pi$ only the $\phi$ absorption events.

The integral in Eq.~(\ref{eq:eikonal1}) does not
depend on the direction of the $\phi$ momentum and thus we
have

\begin{equation}
P_{\textrm{out}}\equiv\frac{\sigma_A}{A\sigma}=\frac{1}{A}\int d^3\vec{r}\rho(r)
\ e\,^{ \int_0^{\infty}
dl \frac{1}{p}\textrm{{\scriptsize Im}}\Pi(p,\rho(r'))}
\label{eq:eikonal2}
\end{equation}

\noindent and this is the magnitude which we would evaluate as a
function of $p$ and $A$, which can be interpreted as the
probability for a $\phi$ to go out of the nucleus.  The density
profiles, $\rho(r)$, for the different nuclei have been taken
from \cite{deVries}.

\section{Evaluation of the $\phi$ selfenergy at finite momentum}

For the evaluation of $\textrm{Im}\Pi$ in nuclear matter we
are going to use the OR and CV models which are based on
chiral $SU(3)$ dynamics considering the $K$ and $\bar K$
in-medium properties. In these works the evaluations were
done for a $\phi$ at rest. Since in the present work we are
dealing with $\phi$ mesons
 with momentum up to $2\textrm{ GeV}$,
  an extension of the models to finite momenta is mandatory.
In the following subsections we summarize these two models
stressing the modifications to consider the finite $\phi$
momentum.

\subsection{Extension of the Oset-Ramos model to finite
momentum
\label{sec:Angels}}

\begin{figure}[h]
\centerline{\protect\hbox{
\psfig{file=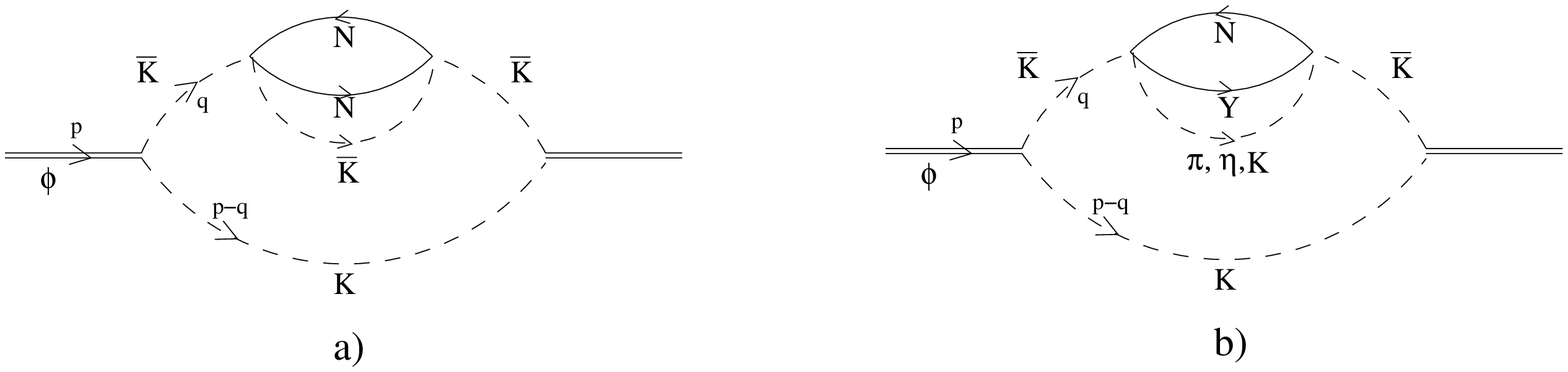,width=0.9\textwidth,silent=}}}
\caption{Diagrams contributing to the $\phi$ selfenergy
coming from the s-wave $\bar{K}N$ interaction.
$Y$ represents $\Lambda$, $\Sigma$ or
$\Xi$.}
\label{fig:diagrams_swave}
\end{figure}

\begin{figure}[h]
\centerline{\protect\hbox{
\psfig{file=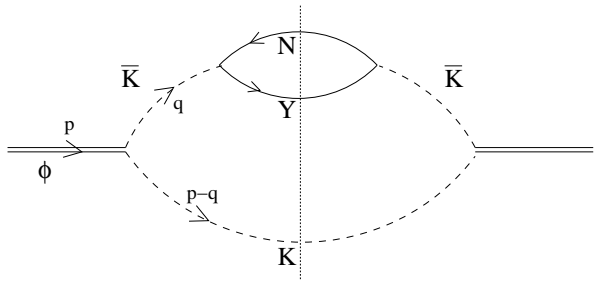,width=0.5\textwidth,silent=}}}
\caption{Diagrams contributing to the $\phi$ selfenergy
coming from the p-wave $\bar{K}N$ interaction.
 $Y$ represents $\Lambda$, $\Sigma$ or $\Sigma^*(1385)$.}
\label{fig:diagrams_pwave}
\end{figure}

The OR accounts for $\phi\to K^+K^-$  and
$\phi\to K^0\bar{K}^0$ decay diagrams where the kaons are
renormalized in the medium due to s-wave and p-wave
interactions, Figs.~\ref{fig:diagrams_swave} and
\ref{fig:diagrams_pwave} respectively.

\begin{figure} [h]
\centerline{\protect\hbox{
\psfig{file=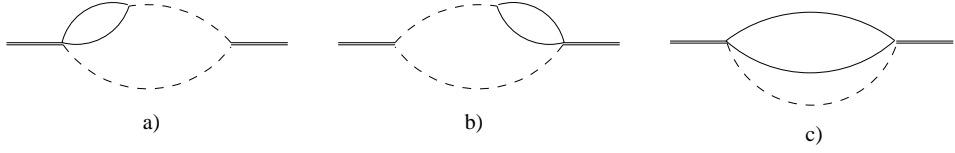,width=0.8\textwidth,silent=}}}
\caption{Diagrams accounting for vertex corrections.}
\label{fig:diagrams_vertex}
\end{figure}

\noindent
To the diagrams of Fig.~\ref{fig:diagrams_pwave}
one must add the vertex corrections shown
in Fig.~\ref{fig:diagrams_vertex} for consistency with
gauge invariance.

In the evaluation of the previous diagrams, the 
$K^+$, $K^0$, selfenergy is
taken as in  \cite{siegel} following the $t\rho$ approximation

\begin{equation}
\Pi_{K^+,K^0}= 0.13\, m_K^2\frac{\rho}{\rho_0}
\textrm{ MeV} \ .
\label{eq:piK+}
\end{equation}

Concerning the $\bar{K}$, a large contribution to the $\phi$
selfenergy comes from the p-wave and vertex correction
diagrams. On the other hand, since the $K^-$ p-wave
selfenergy is proportional to $\vec{q}\,^2$, it is likely to
be more affected inside the loops by the finiteness of the
$\phi$ momentum. Hence we evaluate explicitly the $\phi$
selfenergy from the p-wave $K^-$ selfenergy diagrams at
finite $\phi$ momentum and take the same part coming from the
$K^-$ s-wave selfenergy as obtained for zero momentum. This
latter part produces a contribution to the imaginary part of
the $\phi$ selfenergy that can be parametrized as
\begin{equation}
\label{eq:impiswave}
\textrm{Im}\Pi^{(s)}=
-7.62\rho/\rho_0\,\omega(p)\textrm{ MeV}
\end{equation}

We now evaluate the contribution to $\textrm{Im}\Pi(p,\rho)$ of
the $\bar{K}$ p-wave selfenergy diagrams, depicted in
Fig.~\ref{fig:diagrams_pwave}.

In Fig.~\ref{fig:diagrams_pwave},
 $K$ represents $K^+$ or $K^0$, $\bar K$ represents
$K^-$ or $\bar{K^0}$ and $Y$ the hyperons $\Lambda$,
$\Sigma$ and $\Sigma^*(1385)$.
The $\phi$ selfenergy coming from these diagrams at finite
momentum can be obtained as

\begin{eqnarray} 
\nonumber
-i\Pi^{(p)}(p,\rho)&=2&\int\frac{d^4q}{(2\pi)^4}iD(q)iD(q)
(-i)\Pi^{(p)}_{\bar K}(q,\rho)
i\widetilde{D}(p-q,\rho)\overline{\sum} V_{\phi K\bar K} V_{\phi K\bar K} = \\ 
&=&2g_\phi^2\int\frac{d^4q}{(2\pi)^4}D(q)^2\Pi^{(p)}_{\bar
K}(q,\rho)
\widetilde{D}(p-q,\rho)\frac{4}{3}\left[\frac{(q\cdot p)^2}{m_\phi^2}-q^2\right]
\label{eq:pipwave}
\end{eqnarray}

\noindent 
where $D(q)=(q^2-m_K^2+i\epsilon)^{-1}$ is the $K^-$,
$\bar K^0$,
propagator,  $\widetilde{D}(p-q)=((p-q)^2-m_K^2-\Pi_{K^+})^{-1}$
is the $K^+$, $K^0$,
propagator with the selfenergy of Eq.~(\ref{eq:piK+}). In the
evaluation of Eq.~(\ref{eq:pipwave}) we have used that 
$V_{\phi K\bar K}=-ig_\phi\epsilon_\mu(p_K^\mu-p_{\bar K}^\mu)$,
with $g_\phi=4.57$. The factor $2$ at the beginning of the
equation is due to the possibility of having $K^-K^+$ or 
$\bar K^0 K^0$ in the intermediate states.

To obtain $\textrm{Im}\Pi$ we can apply the following 
Cutkosky rules for the cut represented by the dotted line of
Fig.~\ref{fig:diagrams_pwave}
\begin{eqnarray}
\nonumber \Pi(k)&\quad\to\quad& 2i\,\textrm{Im}\Pi(k)
 \\ \label{eq:cutkosky}
\nonumber D(k)&\quad\to\quad& 2i\,\Theta(k^0)\textrm{Im}D(k)
\end{eqnarray}

\noindent
which leads, after performing the $q^0$ integration, to

\begin{equation}
\textrm{Im}\Pi^{(p)}(p,\rho)=g_\phi^2\int\frac{d^3\vec q}{(2\pi)^3}
\left(\frac{1}{{q^0}^2-\vec{q}\,^2-m_K^2}\right)^2
\frac{1}{\widetilde{\omega}_{p-q}}\textrm{Im}\Pi^{(p)}_{\bar K}(q,\rho)
\frac{4}{3}\left[\frac{(q\cdot p)^2}{m_\phi^2}-q^2\right]
\Theta(q^0)
\label{eq:impipwave}
\end{equation}

\noindent  with 
$\widetilde{\omega}_{p-q}=\sqrt{(\vec p- \vec q)^2
+m_K^2+0.13m_K^2\frac{\rho}{\rho_0}}$, 
$q^0=p^0-\widetilde{\omega}_{p-q}$  and
$\Theta$ is the step function. The $\bar K$ p-wave selfenergy, 
$\textrm{Im}\Pi^{(p)}_{\bar K}(q)$, is evaluated in  \cite{ramos}
and is a function of the $\bar K N Y$ vertices and the Lindhard
functions for hyperons.  In Eq.~(\ref{eq:impipwave}) we also add,
as in  Ref.~\cite{klingl2}, a form factor for each kaon-baryon
vertex of dipole type, $[\Lambda^2/(\Lambda^2-q^2)]^2$, with
$\Lambda=1.05\textrm{ GeV}$.

%
%

Finally, the evaluation of the vertex corrections
 of Fig.~\ref{fig:diagrams_vertex} is
done in an analogous way. In this case we
need the $\phi KNY$ vertex functions given by
\begin{eqnarray} \nonumber
V_{\phi KNY}&=&g_\phi\tilde{V}_{\bar{K}NY}
\,\vec{\sigma}\cdot\vec{\epsilon}\,^{(\phi)};\qquad Y=\Lambda, \Sigma
\\
V_{\phi KN\Sigma^*}&=&g_\phi\tilde{V}_{\bar{K}N\Sigma^*}
\,\vec{S}^{\dagger}\cdot\vec{\epsilon}\,^{(\phi)} \ ,
\end{eqnarray}

\noindent
where $\tilde{V}_{\bar{K}NY}$ and $\tilde{V}_{\bar{K}N\Sigma^*}$
are coefficients given in \cite{oset}. The resulting 
expression for the
imaginary part of the $\phi$ selfenergy coming from the vertex
corrections is the same as Eq.~(\ref{eq:impipwave}) but doing the
following substitution

\begin{equation}
D(q)^2\vec{q}\,^2\frac{4}{3}\left[\frac{(q\cdot p)^2}
{m_\phi^2}-q^2\right] \longrightarrow
D(q)\frac{4}{3}
 \left[ \vec{q}\,^2-
 \frac{(p\cdot q)(\vec{p} \cdot \vec{q})}{m_\phi^2} \right]
+\left(1+\frac{\vec{q}\,^2}{3m_{\phi}^2}\right) \ .
\end{equation}

\subsection{Extension of the Cabrera-Vicente model to finite
momentum
\label{sec:Dani}}

The $\phi$ selfenergy from $\bar{K}K$ loop diagrams, calculated in
\cite{klingl2,dani} for a $\phi$ at rest in a nuclear medium, is given by
\begin{equation}
\label{phi_self}
\Pi_{\phi}^{\bar{K}K} (p^0,\vec{p};\rho) = 2 i
g_{\phi}^2 \frac{4}{3} \int  \frac{d^4q}{(2\pi)^4} 
\bigg \lbrack \frac{(p \cdot q)^2}{m_{\phi}^2}-q^2 \bigg \rbrack
D_K(p^0-q^0,\vec{p}-\vec{q};\rho)
D_{\bar{K}}(q^0,\vec{q};\rho)
\end{equation}
for a $\phi$ meson with a momentum $p$.

A spectral representation
\cite{klingl2} which sums up to all orders the insertion of
irreducible kaon selfenergy terms is used
in the kaon propagator. The imaginary part of
$\Pi_{\phi}^{\bar{K}K}$, which is of interest in the present work, can be
written as
\begin{eqnarray}
\label{Im_phi_self}
\textrm{Im} \Pi_{\phi}^{\bar{K}K} (p^0,\vec{p};\rho)
&=& - \frac{1}{4\pi} g_{\phi}^2 \frac{4}{3} \int_0^{\infty} dq \, \vec{q}\,^2
\int_{-1}^{1} du \frac{1}{\widetilde{\omega}_{p-q}}
\nonumber \\
& &\bigg \lbrack \frac{(p \cdot q)^2}{m_{\phi}^2}-q^2 \bigg
 \rbrack_{q^0=p^0-\widetilde{\omega}_{p-q}}
S_{\bar{K}}(p^0-\widetilde{\omega}_{p-q},\vec{q};\rho) \,
\Theta (p^0-\widetilde{\omega}_{p-q})
\,\,\, ,
\end{eqnarray}
where $u = \vec{p}\cdot\vec{q} / |\vec{p}|\,|\vec{q}|$ and  
$S_{\bar{K} (K)}$ is the spectral function of the $\bar{K} (K)$ meson,
\begin{equation}
\label{spectral_kaons}
S_{\bar{K} (K)}(q^0,\vec{q};\rho) = - \frac{1}{\pi}
\frac{\textrm{Im} \, \Pi_{\bar{K} (K)}(q^0,\vec{q};\rho)}
{| (q^0)^2 - \vec{q}\,^2 - m_K^2 - \Pi_{\bar{K} (K)}(q^0,\vec{q};\rho)|^2}\,\, .
\end{equation}

The main differences with respect to the approach described in section
\ref{sec:Angels} are the following:

$\bullet$ Eq. (\ref{Im_phi_self}) considers higher order effects in the insertion of the
kaon selfenergies as compared to Eqs. (\ref{eq:impiswave}) and
(\ref{eq:impipwave}). Moreover, the calculation at finite momentum is considered
for both the s- and p-wave $\bar{K}$ selfenergy contributions. Note also that
Eq. (\ref{Im_phi_self}) includes the $\phi$ free width into the $\bar{K} K$
channel, which will be subtracted to keep only the nuclear effects in the $\phi$
selfenergy.

$\bullet$ We use the p-wave $\bar{K}$ selfenergy given in \cite{dani} which, starting from
the result in \cite{oset}, takes into account  an improvement of the
relativistic recoil corrections of the $\bar{K} N Y$ vertices. In addition,
these vertices carry a static form factor of the form
$[\Lambda^2/(\Lambda^2+\vec{q}\,^2)]^2$, with $\Lambda=1.05$ GeV.

$\bullet$ The vertex correction diagrams considered in section \ref{sec:Angels} are also
calculated here at finite $\phi$ momentum, with the prescription of using the
fully dressed kaon propagators in the medium.

\section{Results and discussion}

\begin{figure}
\centerline{\protect\hbox{
\psfig{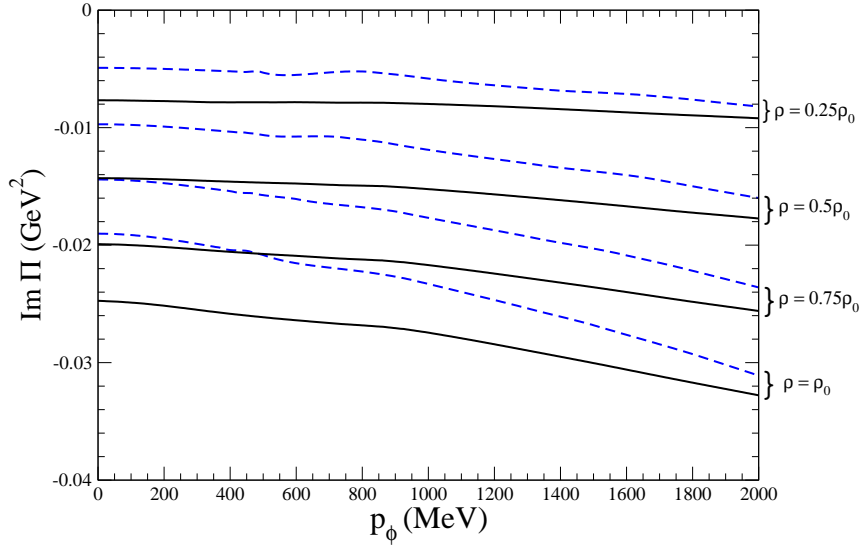}}}
\caption{Imaginary part of the in medium $\phi$ selfenergy,
without the inclusion of the
free part, as a function of the momentum of the
$\phi$ and the density, for the finite momentum modification of
model \cite{oset} (dashed line) and \cite{dani} (solid line). }
\label{fig:ImPi}
\end{figure}

In Fig.~\ref{fig:ImPi} we show the imaginary part of the
$\phi$ selfenergy as a
function of the $\phi$ momentum for different nuclear
densities and for the two models described in
section~\ref{sec:Angels} (dashed line) and
\ref{sec:Dani} (solid line).
The contribution to $\textrm{Im}\Pi$ coming from the
free $\phi$ decay into $K\bar{K}$, non density dependent,
has been subtracted from the full
$\textrm{Im}\Pi$ since the $K\bar{K}$ coming from the
free decay would be
detected and counted as a $\phi$ event, hence it does not
contribute to the loss of flux
required in the argument of the exponential in
Eq.~(\ref{eq:eikonal2}).
Despite the visible
differences in Fig.~\ref{fig:ImPi} between the two models at
zero momentum, the trend of the plot as the momentum
increases is very similar. This initial discrepancy at zero
momentum is expectable due to the differences in the
treatment of the s-wave contribution, the relativistic recoil
corrections and the fact that the model of section
\ref{sec:Dani} goes beyond the first order in density,
implicitly considered in the model of
section~\ref{sec:Angels}.
 The differences between
the two models in  Fig.~\ref{fig:ImPi}, however, are
indicative of the intrinsic theoretical uncertainties.

\begin{figure}
\centerline{\protect\hbox{
\psfig{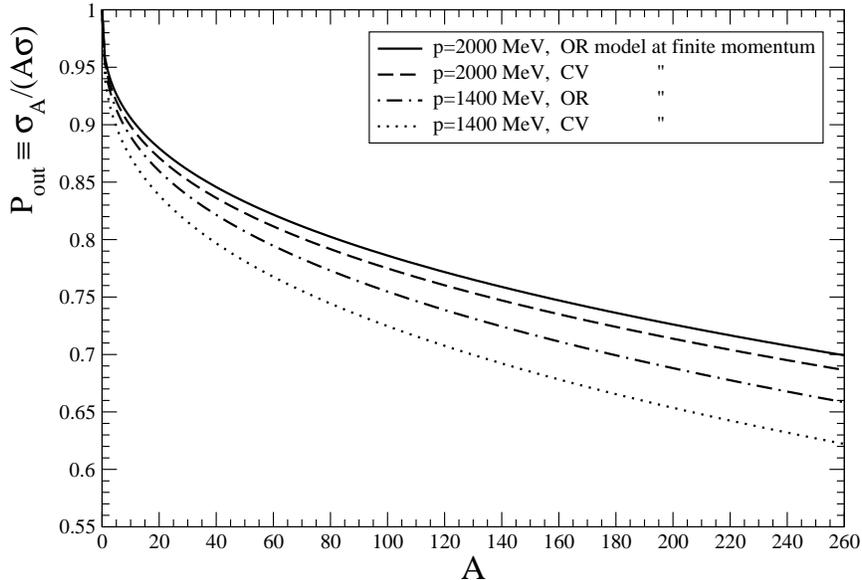}}}
\caption{$\sigma_A/(A\sigma)$ as a function of the nuclear mass
number for two different momentum of the $\phi$ calculated with the
two models described in the text.}
\label{fig:Pout_vs_A}
\end{figure}

In Fig.~\ref{fig:Pout_vs_A} we show the ratio
$\sigma_A/(A\sigma)$, which represents the probability of one
photoproduced $\phi$ to go out the nucleus, calculated
theoretically as a function of $A$ and for two different momenta. 
Again, a comparison between the two models is shown.
We observe that for $p=1400\textrm{ MeV}$ and heavy nuclei this
ratio can be of the order of $0.65$, indicating a clear $\phi$
loss of flux which should be identifiable
experimentally.

\begin{figure}
\centerline{\protect\hbox{
\psfig{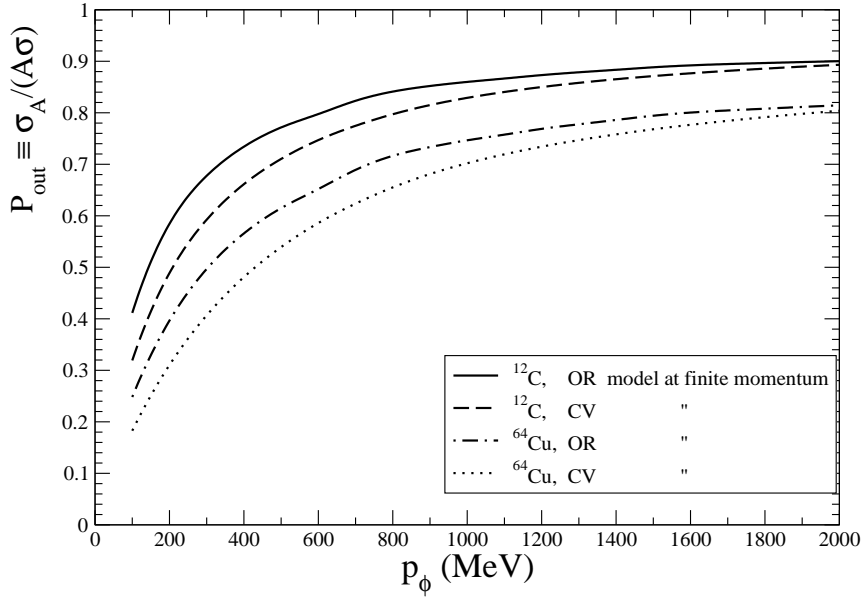}}}
\caption{$\sigma_A/(A\sigma)$ as a function of the
 momentum of the $\phi$ for two different nuclei 
calculated with the
two models described in the text.}
\label{fig:Pout_vs_p}
\end{figure}

In Fig.~\ref{fig:Pout_vs_p} we show instead the results for 
$\sigma_A/(A\sigma)$ as a function of the $\phi$ momentum for two
nuclei and comparing  the two theoretical models. In this figure
we observe that the amount of $\phi$  flux lost is larger for
smaller $\phi$ momentum. This is logical to expect since the
probability of $\phi$ decay per unit length
is $-\textrm{Im}\Pi/p$,
which is larger for small momenta because of the factor $1/p$.
This occurs in spite of the fact that
$-\textrm{Im}\Pi$ decreases with $p$, as can be seen
in Fig.~\ref{fig:ImPi}, because this $p$ dependence is
weaker than that of $1/p$.
 The set up of the experiment at {\it
SPring8/Osaka} \cite{ishikawa} is such that it produces $\phi$
momenta around $1500\textrm{ MeV}$, using photons from $1.4$ to
$2.4\textrm{ GeV}$, coming from $\phi$ production in the forward
direction in the CM frame. A possibility to extract further
information is to use photons of lower energy, around,
$1.6\textrm{ GeV}$, not far from threshold which leads to $\phi$
of around $100\textrm{ MeV}$ in the CM frame and around 
$1000\textrm{ MeV}$ in the lab frame. As one can see in
Fig.~\ref{fig:Pout_vs_p}, in a nucleus like $^{64}{Cu}$ the
depletion factor goes from $0.8$ at $p=2000\textrm{ MeV}$ to
$0.7$ at $p=1000\textrm{ MeV}$. This latter factor would be
$0.57$ for $A\simeq 240$.

So far the only nuclear effects considered have been the ones
related to  the absorption of the $\phi$ meson. At this point
it is worth mentioning that other nuclear effects regarding
the production mechanism may lead to a further $\phi$ loss of
flux or change in the $\phi$ distribution. These other
nuclear effects are mainly the Pauli blocking of the final
nucleon and the Fermi motion of the initial one. The first
one may lead to a reduction of the $\phi$ flux in comparison
to the free case because a certain amount of events are
forbidden due to the Pauli blocking of the final nucleon. On
the other hand the initial Fermi motion can distort the
distribution of the final $\phi$ mesons.   In order to
estimate the possible flux reduction due to these sources, we
have included in the integrand of Eq.~(\ref{eq:eikonal2}) a
factor $G(Q,\rho)$ which considers a Fermi average of these
effects \cite{Garcia-Recio:1987ik}:

\begin{equation}
G(Q,\rho)=1-\Theta(2-\widetilde{Q})\left( 1-\frac{3}{4} 
\widetilde{Q}
+\frac{1}{16}\widetilde{Q}\,^3\right)
\label{eq:Pauli}
\end{equation}

\noindent
where $\widetilde{Q}=|\vec{Q}|/k_F$ with $Q$ the
 momentum transfer
of the nucleon and $k_F=(\frac{3}{2}\pi^2\rho(r))^{1/3}$ is the
Fermi momentum of the nucleons.

\begin{figure}
\centerline{\protect\hbox{
\psfig{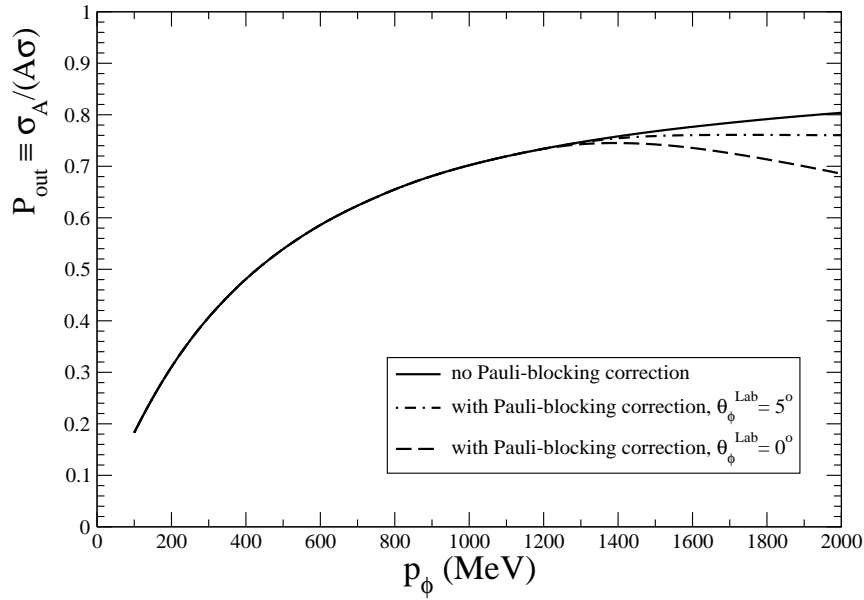}}}
\caption{Effect of the Pauli-blocking correction in the
$\sigma_A/(A\sigma)$ as a function  of the
momentum of the $\phi$ for $^{64}Cu$ calculated with the CV
model.}
\label{fig:Pout_vs_P_Pauli}
\end{figure}

In Fig.~\ref{fig:Pout_vs_P_Pauli} 
we show  $\sigma_A/(A\sigma)$, with the CV model, as a function of
the $\phi$ momentum for $^{64}Cu$. In solid line we show the
result without considering the effect of Eq.~(\ref{eq:Pauli}),
(i.e., the same as in Fig.~\ref{fig:Pout_vs_p}).  The other
lines represent the results considering the Pauli effect
estimation, using Eq.~(\ref{eq:Pauli}). Note that the $Q$
dependence of this $G$ factor introduces another kinematical
variable in the $P_{out}$ observable. We have chosen this
extra variable to be the zenithal angle of the $\phi$ meson in
the Lab frame ($\theta_\phi^{Lab}$). The dashed and
dot-dashed curves
represent the results considering the Pauli effect for
$\theta_\phi^{Lab}=0^o$ and $5^o$ respectively.
We can see that significant effects are obtained for large
$\phi$ momenta, what is expectable since large $\phi$ momenta
imply small final nucleon momenta, which are strongly Pauli
blocked. The Pauli effect also decreases with the $\phi$ meson
angle because the momentum transfer strongly increases with
this angle, therefore the largest effect is obtained for
the $\phi$ forward direction. 
Since the $\phi$ photoproduction is strongly forward peaked,
one can expect $P_{out}$ to be actually somewhere between the
dashed and dot-dashed lines in a real experiment.

\begin{figure}
\centerline{\protect\hbox{
\psfig{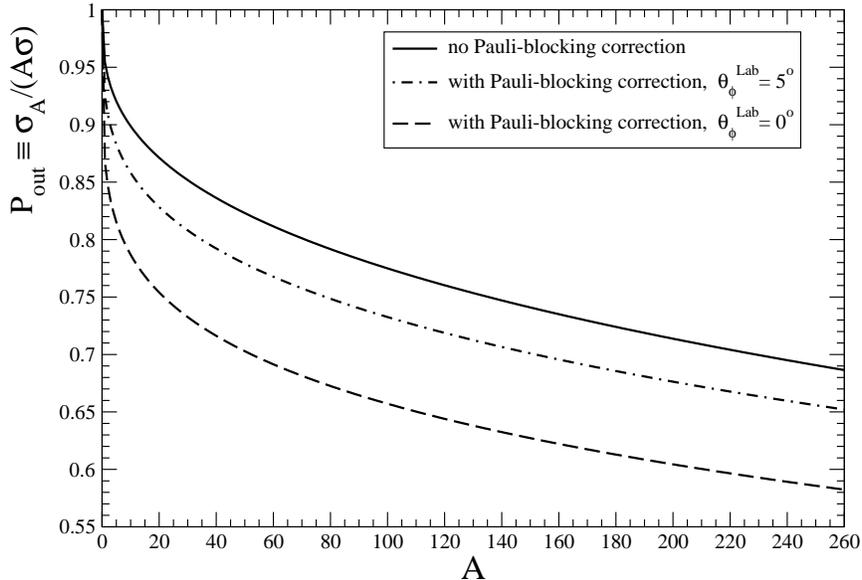}}}
\caption{Effect of the Pauli-blocking correction in the
$\sigma_A/(A\sigma)$ as a function of the
nuclear mass number for $p_\phi=2000\textrm{ MeV}$
calculated with the CV model.}
\label{fig:Pout_vs_A_Pauli}
\end{figure}

In Fig.~\ref{fig:Pout_vs_A_Pauli} we show the $A$ dependence
of $P_{out}$, with the CV model,  without including the Pauli
correction (solid line) and including it for
$\theta_\phi^{Lab}=0^o$ and $5^o$ (dashed and dot-dashed
lines respectively), all for $p_\phi=2000\textrm{ MeV}$ where
the effect is maximum. We observe that the loss of flux due
to the Pauli correction is nearly constant in a wide range of
the mass number, resulting in a reduction of around $0.1$ in
$P_{out}$ for the forward direction. Nonetheless this effect
is relatively smaller for heavier nuclei compared to the
$\phi$ absorption.

   The cross sections calculated in this work are inclusive, 
 summing over all possible final nuclear excited states. The 
 coherent cross section, where the final nucleus is the
 original one,  is not included in the calculation.  Its
 evaluation requires a complete knowledge of the spin and
 isospin dependence of the elementary $\gamma N \to \phi N$
 amplitude and a different treatment than the one done here. 
 The coherent  cross section involves the square of the
 product of the mass number  and the nuclear form factor
 times the square of the spin-isospin averaged amplitude. 
 The  coherent and incoherent cross sections can in principle
 be separated experimentally, as  done in the case of pions
 around the delta energy region \cite{Arends:1982cw}.
 However, this might be difficult at higher energies where
 one usually has a poorer energy resolution.  An alternative
 to this separation, concerning the present work, is to look
 for $\phi$ production at finite angles where the momentum
 transfer to the nucleus is large and the coherent process
 can be neglected. The incoherent cross section sums the
 square of the spin-flip and non spin-flip amplitudes, while
 in the coherent part only the spin independent part
 contributes. Thus, the ratio $|A\,F(Q)|^2/(A\,G(Q))$, where
 $F(Q)$ is the nuclear form factor and $G(Q)$ the Pauli
 blocking factor of Eq.~(\ref{eq:Pauli}), is an upper bound
 for the ratio of coherent to incoherent nuclear cross
 sections. We have studied this ratio as a function of the
 photon energy, the $\phi$ meson angle and the mass number. 
 Our results can be summarized as follows: 
 1) We find that below $E_\gamma =2000 \textrm{ MeV}$ the
 ratio of coherent to incoherent cross section is smaller
 than ten percent for all angles and nuclei beyond $A \sim
 12$. 2) The ratio increases rapidly with the photon energy,
 since the momentum transfer decreases.  For instance at
 $E_\gamma =2400\textrm{ MeV}$, for $^{12}C$ and forward
 angles, the ratio can be of the order of 50 percent, but
 for   heavier nuclei the ratio is smaller.
 For instance, for $^{27}Al$ this ratio is smaller than 10
 percent. 
  We also obtain that beyond 4 degrees the ratio is only a few
 percent and can be safely neglected. In practical terms,
 from the experimental point of view, we can say that in
 nuclei around $^{16}O$ or heavier and photon energies
 smaller than $2000\textrm{ MeV}$, the coherent contribution
 is negligible. This discussion may serve to select an
 experimental setup in which the coherent contribution can be
 neglected, hence facilitating the interpretation of the
 data.


\section{Conclusions}

We  have shown that using present models for the $\phi$
selfenergy in a nuclear medium, conveniently extrapolated to
finite $\phi$ momenta, it is possible to evaluate the
survival rate of $\phi$ produced in nuclear $\phi$
photoproduction, and how this survival rate is tied to the
$\phi$ width in the medium and the momentum of the $\phi$.
The survival rates  for $\phi$ coming from photons in the
range of $1.6$ to $2.4\textrm{ GeV}$  are of the order of
$0.7$, a significant deviation from unity, which are
measurable experimentally. We have shown the $A$ dependence
of the expected $\sigma_A/(A\sigma)$ ratio as well as its
dependence on the $\phi$ momentum. Comparison of the results
with experimental numbers of the incoming experiments would
help determine the accuracy of the models used. These
models could then be used to extrapolate results at other
$\phi$ momenta and one could then get a fair idea of how the
$\phi$ properties are modified in nuclear matter for $\phi$
at any finite momenta below those studied in
 the present work.

\section{Acknowledgments}
D.~C. and L.~R. acknowledge the hospitality of the RCNP of
Osaka University where part of this work was done.
This work is
partly supported by
the Japan-Europe (Spain) Research Cooperation Program of the
Japan Society for the Promotion of Science (JSPS) and the
Spanish Council for Scientific Research (CSIC), by the
 DGICYT contract number BFM2000-1326,
and the E.U. EURIDICE network contract no. HPRN-CT-2002-00311.
D.~C. acknowledges financial support from the Ministerio de
Ciencia y Tecnolog\'{\i}a and L.~R. from the
Ministerio de Educaci\'on, Cultura y Deporte.

\end{document}